\documentclass[fleqn,usenatbib]{mnras}

\usepackage[T1]{fontenc}

\DeclareRobustCommand{\VAN}[3]{#2}
\let\VANthebibliography\thebibliography
\def\thebibliography{\DeclareRobustCommand{\VAN}[3]{##3}\VANthebibliography}

\usepackage{graphicx}	% Including figure files
\usepackage{amsmath}	% Advanced maths commands

\def\be{\begin{equation}}
	\def\ee{\end{equation}}
\def\ba{\begin{eqnarray}}
	\def\ea{\end{eqnarray}}

\newcommand{\dd}{\mathrm{d}}

\newcommand{\bea}{\begin{eqnarray}}
	\newcommand{\eea}{\end{eqnarray}}
\def\beq{\begin{equation}}
	\def\eeq{\end{equation}}

\title[Tracking black holes with popcorn signal]{Tracking the origin of black holes with the stochastic gravitational wave background popcorn signal}
\author[M. Braglia, J. Garc\'ia-Bellido \& S. Kuroyanagi]{
	Matteo Braglia,$^{1,2}$\thanks{E-mail: matteo.braglia@csic.es}
	Juan Garc\'ia-Bellido,$^{1}$\thanks{E-mail: juan.garciabellido@uam.es}
	Sachiko Kuroyanagi$^{1,3}$\thanks{E-mail: sachiko.kuroyanagi@csic.es}
	\\
	$^{1}$Instituto de Fisica Teorica, Universidad Autonoma de Madrid, Madrid, 28049, Spain\\
	$^{2}$INAF/OAS Bologna, via Gobetti 101, I-40129 Bologna, Italy\\
	$^{3}$Department of Physics and Astrophysics, Nagoya University, Nagoya, 464-8602, Japan
}

\date{Accepted XXX. Received YYY; in original form ZZZ}

\pubyear{2021}

\begin{document}
\label{firstpage}
\pagerange{\pageref{firstpage}--\pageref{lastpage}}
\maketitle

\begin{abstract}
Unresolved sources of gravitational waves (GWs) produced by the merger of a binary of black holes at cosmological distances combine into a stochastic background. Such a background is in the continuous or popcorn regime, depending on whether the GW rate is high enough so that two or more events overlap in the same frequency band. These two regimes respectively correspond to large and small values of the so-called {\it duty cycle}. We study the detection regime of the background in models of Primordial Black Holes (PBHs) and compare it to the one produced by black holes of stellar origin. Focusing on ground-based detectors, we show that the duty cycle of the PBH-origin background is larger than that of astrophysical black holes because of differences in their mass function and the merger rate. Our study opens up the possibility to learn about the primordial or astrophysical nature of black hole populations by examining the statistical properties of the stochastic background. 
\end{abstract}


%%%%%%%%%%%%%%%%%%%%%%%%%%%%%%%%%%%%%%%%%%%%%%%%%%

%%%%%%%%%%%%%%%%% BODY OF PAPER %%%%%%%%%%%%%%%%%%

\section{Introduction}

The Stochastic Gravitational Wave Background (SGWB) is one of the most interesting targets of current and future gravitational wave (GW) observatories. A detection of a SGWB from the Early Universe would provide a breakthrough in our understanding of the origin of the Universe. In addition to those of primordial origin, we expect SGWBs composed of many astrophysical sources at cosmological distances that cannot be individually resolved~\citep{Regimbau:2011rp,Rosado:2011kv}. Example of sources contributing to such an astrophysical background are core-collapse supernovae~\citep{Ferrari:1998ut}, neutron stars (both during their formation~\citep{Coward:2000fe} and rotation~\citep{Ferrari:1998jf,Regimbau:2001kx}), magnetars~\citep{Regimbau:2005ey,Chowdhury:2021vqn}, the early inspiral phase of compact binaries~\citep{Farmer:2003pa,Regimbau:2005tv} or the coalescence of very massive  BHs~\citep{Sesana:2004gf,Sesana:2008mz}. Such astrophysical backgrounds can be used as a tool to constrain the properties of such sources. Moreover, each of these signals is associated with a characteristic frequency and spectral shape, 
which can help infer which astrophysical sources are contributing to the background.

Besides the frequency dependence, there are other properties that distinguish the nature of different backgrounds, such as their anisotropies~\citep{Cusin:2018rsq,Jenkins:2018uac,Jenkins:2018kxc,Cusin:2019jpv,Bertacca:2019fnt,Pitrou:2019rjz}, polarizations~\citep{Cusin:2018avf}, and popcorn (sometimes referred as non-Gaussian or time-dependent) signal~\citep{Coward:2006df,Regimbau:2008nj,Wu:2011ac,Mukherjee:2019oma}. Those characteristics would be essential for disentangling different possible sources and identifying the origin of the SGWB. The popcorn signal, which is the focus of this paper, could be seen in a SGWB originating from overlapped GW souces~\citep{Coward:2006df,Regimbau:2011bm}. It depends on the relative duration of the transient signal and the time interval between successive events. If such interval is small compared to the duration of a single event and/or the number of sources is very large, the GW events overlap, and the background is in the so-called {\em continuous} regime, characterized by Gaussian statistics, as a consequence of the central limit theorem. On the other hand, if the interval between events is comparable or larger than the typical duration of the signal, the waveforms may or may not overlap, and the statistical properties are strongly non-Gaussian. We denote this regime as {\em popcorn} background. A useful quantity to distinguish between the two regimes is the so-called astrophysical {\em duty cycle}, which represents the average number of events present in a given frequency band. Continuous and popcorn regimes correspond to large and small duty cycles, respectively (i.e. values of the duty cycle larger or smaller than one).

Our goal in this paper is to show that the popcorn signature of the SGWB can be used to learn about the nature of  BHs. The cumulative detection of GWs from binary black hole (BBH) inspirals has attracted wide attention to Primordial Black Hole (PBH) as a possible source~\citep{Bird:2016dcv,Sasaki:2016jop,Clesse:2016vqa}. Contrary to standard astrophysical black holes (ABHs), which form by the collapse of massive stars at the end of their life cycle, PBHs could have been formed by the collapse of very large density perturbations during the radiation era~\citep{1967SvA....10..602Z,Hawking:1971ei}. The amplification of curvature perturbations produces the seeds of such perturbations during inflation at scales much smaller than those tested by Cosmic Microwave Background (CMB) observations~\citep{Carr:1993aq,Carr:1994ar,Garcia-Bellido:1996mdl}. Being almost collisionless and {\em dark} by definition, PBHs cosmologically behave as Cold Dark Matter (CDM) and make up a fraction of it. They can take masses well below Chandrasekhar mass, with the only constraint being that PBHs lighter than $10^{ 15}$g would have already evaporated by today. Besides their mass function, PBHs also differ in their merger rate, which is expected to grow with redshift, unlike that of ABHs, which follow the star formation rate~\citep{Cholis:2016xvo,Raidal:2017mfl,Vaskonen:2019jpv,Atal:2020igj,DeLuca:2020qqa,Mukherjee:2021ags}. 

The observation of BBHs indicates that the SGWB from their superposition could be detected in the near future by an upgraded ground-based detector network~\citep{LIGOScientific:2016fpe,KAGRA:2021kbb,Mandic:2016lcn,Clesse:2016ajp,Wang:2016ana}. Once detected, one of the primary challenges is to identify whether its origin is astrophysical or primordial\footnote{In this paper, we consider the SGWB produced by PBH binaries. We note that PBHs source a SGWB at their formation through second order coupling of tensor and scalar cosmological fluctuations~\citep{Acquaviva:2002ud,Saito:2008jc,Saito:2009jt}. At the frequencies tested by ground-based interferometers, the background is produced by the formation of very small mass PBHs which have already evaporated. LIGO/Virgo data have been used recently to constrain the abundance of such PBHs~\citep{Kapadia:2020pnr,Romero-Rodriguez:2021aws}. For the PBH masses considered in this paper, such $\emph{scalar-induced}$ background cannot be tested with ground-based interferometers. It mainly contributes at very small GW frequencies and in fact has been proposed in~\citep{Vaskonen:2020lbd,DeLuca:2020agl,Kohri:2020qqd} as an explanation to the possible GW signal recently indicated by NANOGrav~\citep{NANOGrav:2020bcs}. }. In this paper, we explore the possibility of using the duty cycle for distinguishing BBH formation mechanisms. One approach to address this issue is to use the spectral shape~\citep{Mukherjee:2021ags,Bavera:2021wmw}, while this is the first study where the SGWB from PBHs is studied using the duty cycle and going beyond its spectral shape. Focusing on ground-based interferometers, we find that the duty cycle for PBH is generically higher than the one of astrophysical populations. We develop a simple procedure to compute the duty cycle, taking into account the sensitivity of the detectors. This allows selecting, from all the events contributing to the duty cycle, only those seen by a given detector with a Signal-to-Noise-Ratio (SNR) exceeding a certain threshold. Our results are suitable to be used in population searches with unresolved events \citep{Smith:2020lkj,Biscoveanu:2020gds}.

Our paper is organized as follows. In Sec.~\ref{sec:mass_function}, we describe the PBH models that we use in our work. We consider not only the widely used log-normal mass function, but also the mass function motivated by the thermal history of the universe. The theoretical framework is reviewed in Secs.~\ref{sec:merger_rate}, \ref{sec:SGWB} and \ref{sec:duty_cycle} which are dedicated to the merger rate, the SGWB and the duty cycle respectively. We finally present our results in Sec.~\ref{sec:results} and conclude in Sec.~\ref{sec:conclusions}.

\section{Primordial Black Hole models}
\label{sec:mass_function}

\begin{figure}
	\includegraphics[width=\columnwidth]{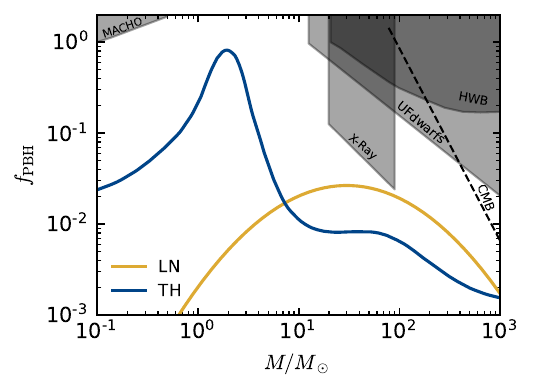}
	\caption{ \label{fig:fpbh} The Lognormal (LN) and Thermal History (TH)  PBH mass functions are shown in yellow and blue respectively. The shaded regions are constraints from 
	microlensing (MACHO), 
	ultra-faint dwarf galaxies and Eridanus II (UDFdwarfs)~\citep{DES:2016vji}, 
	X-ray/radio counts (X-Ray)~\citep{Gaggero:2016dpq}, see however~\citep{Scarcella:2021jzp}, and
	halo wide binaries (HWB)~\citep{Quinn:2009zg}. 
	The accretion constraint (CMB)~\citep{Ali-Haimoud:2016mbv,Poulin:2017bwe,Serpico:2020ehh} is shown with the dashed line because the modeling of accretion adopted enters a different regime at the large masses ($\sim10^4\,M_\odot$)~\citep{Clesse:2017bsw,Carr:2019kxo}. Note that the constraints are obtained for the monochromatic mass function and can change for broad mass function~\citep{Carr:2017jsz,Garcia-Bellido:2017xvr,Calcino:2018mwh}. }
\end{figure}
PBHs form when the overdensities produced during inflation are larger than a critical threshold $\delta_{c}\equiv \delta \rho / \rho$ at horizon re-entry. The latter is sensitive to the energy content of the cosmic fluid and its Equation of State (EOS)~\citep{Musco:2012au}, which is usually taken to be $w=1/3$ during radiation dominated era. The fraction of horizon patches that collapse into PBHs is given by~\citep{Carr:1975qj}
\begin{equation}
	\beta( m )
	\approx
	{\rm erfc}\!
	\left[
	\frac{\delta_{c}\big[ w( m  ) \big]}
	{ \sqrt{2} \, \delta_{\rm rms}( m )}
	\right]
	,
	\label{eq:beta}
\end{equation}
where `erfc' is the complementary error function. Note that we have explicitly expressed the dependence of the critical density in terms of $m$, the mass of the formed PBHs. 

In Eq.~\eqref{eq:beta}, the root-mean-squared amplitude of density perturbations $\delta_{\rm rms}$ is given in terms of the primordial power spectrum of the curvature perturbations $\mathcal{P}_\zeta(k)$ produced during inflation as:
\begin{equation}
	\delta_{\rm rms}^2[ m(k) ]=\int \dd\ln k\,W^2(k R) \left(kR\frac{2}{3}\right)^4\mathcal{P}_\zeta(k)
\end{equation}
where, for simplicity, we have assumed that curvature perturbations from inflation are Gaussian distributed, and $W$ is a window function smoothing over a comoving scale $R\simeq2 G m/a_{\rm form}\gamma^{-1}$. Here, $\gamma$ parametrizes the ratio between the PBH mass and the mass of the collapsing horizon-sized region at PBH formation, which typically takes values $\gamma\in[0.1,\,1]$. Note that the scale of perturbations $k$ correspond to the mass of PBHs as 
\begin{equation}
		\label{eq:mass}
	m(k)\sim 30 M_\odot\left(\frac{\gamma}{0.2}\right)\left(\frac{g_*}{10.75}\right)^{-1/6}\left(\frac{k}{2.9\times10^5\, {\rm Mpc}^{-1}}\right)^{-2}
\end{equation}
where $g_*(T)$ is the number of relativistic degrees of freedom at formation.

From Eq.~\eqref{eq:beta}, we can compute the mass fraction of PBHs per logarithmic interval of masses as
\begin{equation}
	f_{\rm PBH}( m ) 
	\equiv \frac{\Omega_{\rm PBH}(M)}{\Omega_{\rm CDM}}
	\approx 2\left( 1 + \frac{\Omega_b}{  \Omega_{\rm CDM}}\right)\,\beta( m ) \sqrt{\frac{M_{\rm eq}}{m}}
	\, ,
	\label{eq:fPBH}
\end{equation}
where $\Omega_{\rm CDM}=0.245$ and $\Omega_b=0.0456$ are CDM and baryon
density parameters, and $M_{\rm eq} = 2.8 \times 10^{17}\,M_\odot$ is the horizon mass at matter-radiation equality. The total fraction of PBHs is obtained by integrating $f_{\rm PBH}( m )$ over its full support
\begin{equation}
	\label{eq:ftot}
	f_{\rm PBH}^{\rm tot}\equiv\int\,d\ln m \,f_{\rm PBH}( m )\,,
\end{equation}
which should satisfy $f_{\rm PBH}^{\rm tot}\leq1$.

As clear from Eq.~\eqref{eq:beta}, the PBH mass function is affected by the shape of the primordial power spectrum and the EOS parameter\footnote{ Non-Gaussianities of primordial perturbations also affect the PBH mass function. In this paper, for simplicity, we restrict to Gaussian perturbations. For the effects of non-Gaussianities, see e.g.~\citep{Young:2013oia,Garcia-Bellido:2017aan,Franciolini:2018vbk,Atal:2018neu,DeLuca:2019qsy,Yoo:2019pma,Ezquiaga:2019ftu,Kitajima:2021fpq}.}. In this paper, we consider two different PBH models. One is the widely used mass function motivated by inflationary models producing a peaky primordial power spectrum. The other is the mass function that carries the effect of the changes of $w$ due to the thermal history evolution. 

As a first example, we consider the {\bf Lognormal (LN)} mass function~\citep{Dolgov:1992pu}
\begin{equation}
	\label{eq:lognormal}
	f_{\rm PBH}(m)=\frac{f_{\rm PBH}^{\rm tot}}{\sqrt{2 \pi}\sigma}\exp\left[-\frac{\ln^2 m/\mu}{2\sigma^2}\right] \,,
\end{equation}
which is the most widely considered mass function and arises in many inflationary models featuring a peak in the power spectrum~\citep{Clesse:2015wea,Braglia:2020eai}. As a reference, we adopt the values $\mu=30\,M_\odot$ and $\sigma=1.5$. In order to avoid astrophysical constraints, we assume $f_{\rm PBH}^{\rm tot}=0.1$. 

We refer to the second model as {\bf Thermal History (TH)} model. This mass function has a rich structure induced by the thermal history evolution. As the temperature of the Universe decreases, Standard Model particles become non-relativistic when the temperature becomes comparable to the mass scale of the particles. This and the QCD phase transition cause small drops in $w$ and induce pronounced features in the PBH mass function as a result of the exponential dependence of the PBH formation probability on the critical threshold $\delta_c(w)$~\citep{Jedamzik:1996mr,Byrnes:2018clq,Carr:2019kxo}. For the primordial power spectrum, we assume a nearly scale-invariant spectrum of the form $\mathcal{P}_{\zeta}(k)=A_s (k/k_{\rm QCD})^{n_s-1}$ at small scales. Here $k_{\rm QCD}$ is the pivot scale and is the mode that re-enters the horizon at the QCD transition. Note that $A_s$ and $n_s$ take values that are independent of the ones measured at CMB scales. This spectrum is a good approximation to the one produced in inflationary models featuring the second stage of slow-roll with first slow-roll parameter $\epsilon_{\rm QCD}\ll\epsilon_{\rm CMB}$ \citep{Garcia-Bellido:2017mdw,Ezquiaga:2017fvi}. As a reference, we take $n_s=0.97$, and $A_s$ is fixed to satisfy the condition of $f_{\rm PBH}^{\rm tot}=1$.

The mass functions of the two models are shown in Fig.~\ref{fig:fpbh}. As can be seen, the TH mass function shows two pronounced peaks at $m\sim 2\,M_\odot$ and $\sim 70\, M_\odot$. This is the result of taking into account the variations of the equation of state, which deviates from $w=1/3$ at the QCD transition in the Thermal Model of particle physics\footnote{Phase transitions around the QCD epoch and/or Lepton Flavour Asymmetries also modify the PBH mass function~\citep{Bodeker:2020stj,Garcia-Bellido:2021zgu}. We have tested that the results in this paper are qualitatively not very different from the TH case.} and when pions become non-relativistic. Although the plot in Fig.~\ref{fig:fpbh} is restricted to PBH masses of interest for ground-based detectors, we note that the mass function shows secondary peaks also at $m\sim 10^{-5}\,M_\odot$ and $\sim 10^{6}\, M_\odot$. 

Note that functional form of the LN mass function in Eq.~\ref{eq:lognormal} would be also modulated by the variations of $w$ at the QCD transition and the pion mass scale~\citep{Carr:2018poi}. However, being the mass function very peaked, our results are almost unaffected by this. 

Taken the CMB constraints (dashed line in Fig.~\ref{fig:fpbh}) at face value, the presence of the secondary peak around $\sim 10^{6}\, M_\odot$ would rule out the TH mass function. We stress, however, that the results presented in this paper are mainly affected by the masses in the range $[\mathcal{O}(0.1)\,M_\odot,\,\mathcal{O}(100)\,M_\odot\,]$ because of the limited sensitivity and frequency range of the ground-based experiments, and larger masses are almost irrelevant for our purposes. We also note that a possible way to avoid such constraints is to add a small amount of running to the spectral index in the simple parameterization for $\mathcal{P}_{\zeta}(k)$\footnote{Private communication from one of the authors (JGB) to the author of~\citet{Hasinger:2020ptw}.}, which is well motivated since it mimics the rise of the power spectrum from the large scales probed by the CMB~\citep{Ezquiaga:2017fvi}. 
Furthermore, the constraints shown in Fig. 1 have multiple caveats, some due to the assumption of single-mass (so-called monochromatic) mass functions or that of uniformly distributed PBH. Relaxing those assumptions, many of the constraints either go away of they shift around. A reanalysis of some of the bounds can be found in~\citep{Carr:2019kxo}.

\begin{figure*}	
	\includegraphics[width=\columnwidth]{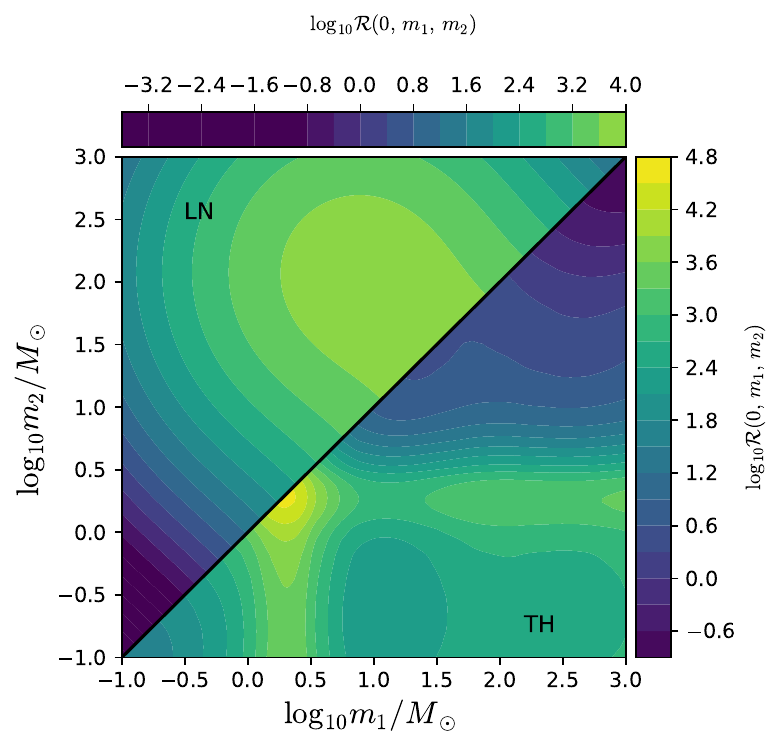}
	\includegraphics[width=\columnwidth]{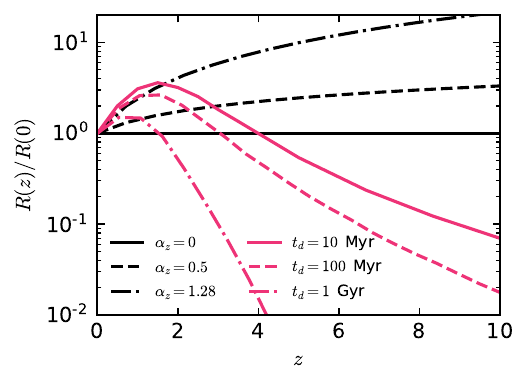}
	\caption{\label{fig:tau} [Left] Local merger rate per comoving volume and logarithmic interval of masses for the LN (upper half) and TH (lower half) models.	[Right] Redshift evolution of the PBH merger rate (black) compared to that of astrophysical ones.  }
\end{figure*}

\section{Merger rates}
\label{sec:merger_rate}

The next step toward a computation of the SGWB is to build a model for the merger rate of the black hole binaries. 

{\bf Primordial Black Holes.} For PBHs, two formation channels exist, both in principle contributing to the total merger rate. In the first channel, so-called early binary, binaries are formed during the radiation era by the tidal torques from other PBHs~\citep{Nakamura:1997sm,Sasaki:2016jop}.  

The other channel, called late binary, PBH binaries are created by tidal capture of PBHs in dense halos~\citep{Quinlan:1989,Mouri:2002mc,Clesse:2016vqa}.  It is still unclear which gives the dominant contribution; especially the PBH clustering is important as it suppresses the former and enhances the latter~\citep{Raidal:2017mfl,Bagui:2021dqi}. In this paper, we consider the late binary formation channel because the merger rate of early binaries is estimated under the assumption that the PBH mass function does not extend over many orders of magnitude~\citep{Kocsis:2017yty,Gow:2019pok}, and the application to a broad mass function, such as the one of the TH model, is not clearly understood. We quickly comment on that in the Conclusions~\ref{sec:conclusions}.

We assume that the differential merger rate per unit time, comoving volume, and mass interval takes the following form:
\begin{equation}
	\label{eq:tauz}
	\frac{\dd^2\tau_{\rm merg}(z,\,m_1,\,m_2)}{\dd  \log_{10}m_1\,\dd \log_{10} m_2}=\mathcal{R}(0,\,m_1,\,m_2)\,(1+z)^{\alpha_z}\,{\rm yr}^{-1} {\rm Gpc}^{-3}\,,
\end{equation}
where, in the case of late binaries, $\mathcal{R}$ is given explicitly by the following expression  \citep{Clesse:2020ghq}
\begin{equation}
    \label{eq:tau0} 
	\mathcal{R}(0,\,m_1,\,m_2)=R_{\rm clust}\,f_{\rm PBH}(m_1)f_{\rm PBH}(m_2)
	\frac{(m_1 + m_2)^{10/7} }{(m_1 m_2)^{5/7}}\,,
\end{equation}
and the total merger rate is the integral of Eq.~\eqref{eq:tauz} over $\dd z\,\dd\log_{10} m_1\,\dd\log_{10} m_2$. The clustering dynamics of PBHs and the time dependence of the merger rate are not clearly understood, and we parameterize them with the two constants $R_{\rm clust}$ and $\alpha_z$ \citep{Clesse:2020ghq, Mukherjee:2021ags,Mukherjee:2021itf}~(see also~\citet{Atal:2022zux} for a different parameterization of the merger rate redshift dependence).

In practice, we use $R_{\rm clust}$ to normalize the total merger rate to the value of $45\,{\rm yr}^{-1}\,{\rm Gpc}^{-3}$, consistently with the upper bound of the 90\% credible interval on the local merger rate inferred from GWTC-3~\citep{LIGOScientific:2021psn}. This amounts to the assumption that all the observed binary  BH events during the second observation run of LIGO-Virgo consist of PBH binaries. It requires the value of $R_{\rm clust}$ to be of the order of $10^3-10^5$, depending on the specific value of $n_s$~\citep{Braglia:2021wwa}. Such large values of $R_{\rm clust}$ are expected if PBHs are strongly clustered~\citep{Clesse:2020ghq}. 

The mass distribution of the merger rate for our PBH models is shown in the left panel of Fig.~\ref{fig:tau}. We see that the two models show quite different features. For the LN case, it peaks at the mass $\mu$, and quickly decays moving away from it. For the TH case, the merger rate is relatively large even away from the peaks induced by the thermal history of the Universe. Indeed, since Eq.~\eqref{eq:tau0} increases as the mass ratio $q=m_2/m_1$ gets small, we expect many events involving secondary masses $m_2< 3\,M_\odot$. The fact that many of such events are predicted in the TH model has important implications for the computation of the duty cycle in the next Sections.

Since the merger rate of PBHs is expected to grow with redshift, we take $\alpha_z$ to be a positive index. As a reference value, we use $\alpha_z=0$, corresponding to a constant merger rate as assumed in~\citep{Clesse:2016vqa,Clesse:2016ajp,Clesse:2020ghq}. We note that, besides $\alpha_z=0$, another typical value often assumed in the literature is $\alpha_z=1.3$~\citep{Raidal:2017mfl,Raidal:2018bbj}. 

{\bf Astrophysical Black Holes.} 
 The phenomenological expression for the merger rate is~\citep{Safarzadeh:2020qru,Mukherjee:2021ags}:
\begin{align}	
    		\frac{\dd^2\tau_{\rm merg}(z_m,\,m_1,\,m_2)}{\dd m_1\,\dd m_2}=\mathcal{N}\,P(m_1,\,m_2)\notag\\\label{eq:mergerastro}\times\int_{z_m}^{\infty}\,\dd z_f\frac{\dd t_f}{\dd z_f}
		\frac{1}{t_d(z_m,\,z_f)}\frac{(1+z_f)^{2.7}}{1+\left(\frac{1+z_f}{2.9}\right)^{5.6}}.
\end{align} 
In the equation above, $\mathcal{N}$ is a normalization constant and we fix it in the same way as PBHs by requiring that $\frac{\dd^2\tau_{\rm merg}(z_m,\,m_1,\,m_2)}{\dd m_1\,\dd m_2}$ integrated over the component masses and evaluated at $z=0$ gives a total rate of $45\,{\rm yr}^{-1}\,{\rm Gpc}^{-3}$.  $z_f$ and $z_m$ denote the reshift at which the binaries form and merge respectively and we use the Madau-Dickinson relation~\citep{Madau:2014bja} for the star-formation rate $R_{\rm SFR}(z)=(1+z)^{2.7}/\left[1+\left(\frac{1+z}{2.9}\right)^{5.6}\right]$. Finally, $t_d(z_m,\,z_f)\equiv t(z_m)-t(z_f)$ is the time delay between formation and merger.  As can be seen from the right panel of Fig.~\ref{fig:tau}, the time-delay governs the redshift evolution of the merger rate. Since models of stellar population synthesis predict different values of $t_d$ that can range from hundreds of Myr up to the age of the Universe, the time-delay is the main source of uncertainty in the computation of the merger rate~\citep{Safarzadeh:2020qru}. Following~\citep{Mukherjee:2021ags}, we choose a probability distribution of the time delay that scales as $1/t_d$. The joint population distribution on $m_1$ and
$m_2$ follows $P(m_1,\,m_2)\propto \Theta(m_1-m_2)\,m_1^{-2.3}/(m_1-M_{\rm min})$, where the Heaviside theta ensures that the primary mass is always larger than the secondary one~\citep{LIGOScientific:2016dsl,LIGOScientific:2019vic}. The normalization is fixed so that the integral of $P$ over ${\rm d}m_1\,{\rm d}m_2$ in the range $[M_{\rm min}=5\,M_\odot,\,M_{\rm max}=50\,M_\odot]$ is 1  (for more details see e.g.~Appendix~D of~\citep{LIGOScientific:2016dsl}).

In the right panel of Fig.~\ref{fig:tau}, we compare the redshift dependence of the PBH merger rate for different values of $\alpha_z$ and that of ABHs. The latter depends on the time delay $t_{\rm d}$ between the binary formation and merger, which, as stressed above, is poorly understood. Nevertheless, independently on $t_{\rm d}$, it is clear that its evolution follows the star-formation rate and peaks around $z\sim2$, unlike the merger rate of PBHs which does not decay for larger redshift.

\section{Stochastic Gravitational Wave Background}
\label{sec:SGWB}

We now discuss the SGWB generated by BH mergers. The energy density of GWs is expressed as an integral over redshift and  masses as follows:
\begin{align}
	\Omega_\text{\tiny GW} (f,\,z_{\rm min},\,z_{\rm max} )=& \frac{f}{\rho_c} \int_{z_{\rm min}}^{z_{\rm maz} }\dd z'\, \dd m_1 \,\dd m_2 \,\frac{1}{(1+z')H(z')}\notag\\&\times  \frac{\dd^2 \tau_{\rm merg}(z',\,m_1,\,m_2)}{\dd m_1 \dd m_2}  \frac{\dd E_\text{\tiny GW} (f_s)}{\dd f_s},
	\label{Eq:OGW}
\end{align}
where $f_s$ is the redshifted source frame frequency $f_s =f (1+z)$, $\rho_c = 3 H_0^2/8\pi G$ is the critical energy density of the Universe. We note that the mass integral is performed over the mass function support, while the redshift integral is between $z_{\rm min}=0$ and $z_{\rm max}$ defined below. Note that, by setting $z=0$, we keep all the binaries in the computation of $\Omega_{\rm GW}$, including also those detectable as single events. 

The single source energy spectrum is given by the following expression in the non-spinning limit~\citep{Ajith:2009bn} 
\begin{equation}
	\frac{\dd E_\text{\tiny GW} (f)}{\dd f} =  \frac{\pi^{2/3}}{3} (G\mathcal{M}_c)^{5/3} f^{-1/3} \mathcal{G}(f)
\end{equation}
where the function $\mathcal{G}(f)$ describes the frequency dependence during the inspiral, merger, and ringdown phases (see e.g. Eqs.~(3.2)-(3.4) of~\citep{Braglia:2021wwa}).
Since the function $\mathcal{G}(f)$ has a cutoff at the maximal emission frequency of the ringdown phase ($f_3$ in the notation of~\citep{Braglia:2021wwa}), the upper limit of the redshift integration in Eq.~\eqref{Eq:OGW} is given by $z_{\rm max}=\frac{f_3}{f}-1$.

\begin{figure}
	\includegraphics[width=\columnwidth]{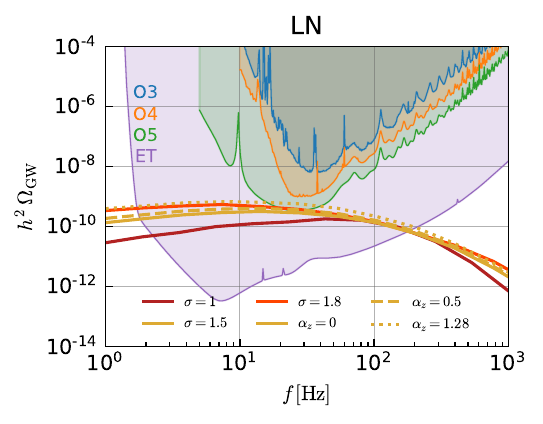}
	\includegraphics[width=\columnwidth]{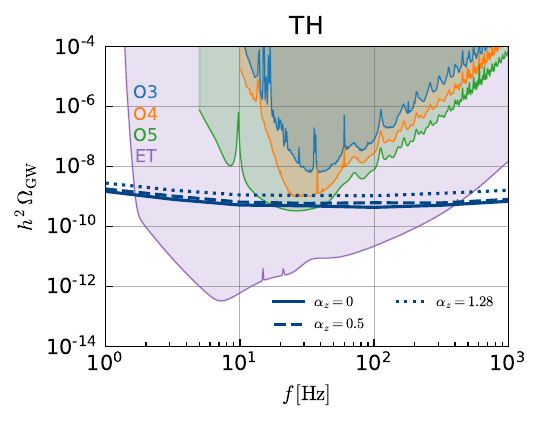}	\includegraphics[width=\columnwidth]{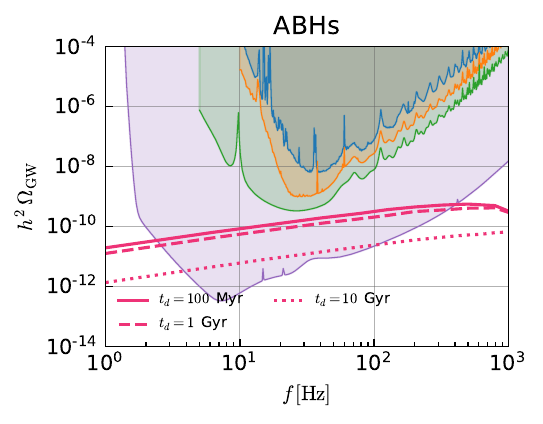}
	\caption{\label{fig:omega} The spectral energy density of the SGWB $\Omega_{\rm GW}$ from PBH binaries for different  BH populations. [Top]  LN model. [Center] TH model. In both panels, solid, dashed, and dotted lines correspond to $\alpha_z=0,\,0.5,\,1.28$, respectively. In the LN case, we also vary the width of the mass function $\sigma$. [Bottom] SGWB of astrophysical origin, for which we vary the time-delay parameter $t_d$.}
\end{figure}

In Fig.~\ref{fig:omega}, we plot some examples of the SGWB spectra from PBH binaries, together with the noise curves of LIGO O3, O4, O5, and Einstein Telescope (ET). The sensitivity curves for the cross-correlation analysis have been computed using the threshold of ${\rm SNR}=1$ and a frequency resolution of $\Delta f /f=0.1$. We have taken the observation time to be $1\,{\rm year}$, except for O3 for which we have chosen the actual observation time described in \citep{KAGRA:2021kbb}. For comparison, we also plot the SGWB produced by ABHs, calculated by Eq.~\eqref{Eq:OGW} using the merger rate in Eq.~\eqref{eq:mergerastro}. We see that different population models show  different amplitude and spectral shapes of the SGWB. 

The amplitude of the SGWB in PBH models is generically larger than that of ABHs if they are to explain all of the LVK observed events\footnote{The possibility of having mixed populations of PBHs and ABHs, as considered in e.g.~\citep{Mukherjee:2021ags,DeLuca:2021wjr,Franciolini:2021tla}, would also be very interesting, but for simplicity, we do not consider it here. In that case, also the contribution from PBH-ABH mergers should be taken into account~\citep{Kritos:2020wcl,Cui:2021hlu}}. Both the LN and TH models have similar amplitudes at the peak sensitivity of LIGO, but their spectral shapes are significantly different. This will be crucial for ET, which has a broader frequency sensitivity, to discriminate between these models (and ABHs). The reason for the different spectral shapes is that the TH population contains binaries with a broad range of total masses and mass ratios. The cutoff in the GW energy spectrum is roughly determined by the total mass of the merger. 
Since the mass function of the TH model contains 
PBHs with very small masses,
this cutoff is pushed to very high frequencies outside the plot. Similarly, for the LN case, we observe a broadening of the peak for larger values of $\sigma$. 

The redshift dependence of the merger rate affects mainly the amplitude of $\Omega_{\rm GW}$ and does not significantly alter its spectral shape. Such redshift-dependent effects are seen in the plots for all the population models. For PBHs, an increasing $\alpha_z$ increases the amplitude of the SGWB. On the other hand, a larger time delay between the formation and merger of astrophysical binaries implies fewer events that contribute to the background, which is therefore characterized by a smaller amplitude.

\section{Duty cycle}
\label{sec:duty_cycle}

\subsection{Standard definition}

Another property of the SGWB, besides its amplitude and spectral dependence, is its statistical behavior. 
The duty cycle is one possible observable to characterize the non-Gaussianity of the SGWB. It is defined by the ratio of the duration of the signal emitted between frequency $f$ and $f+\dd f$ and the time interval between two events \citep{Coward:2006df}:
\begin{equation}
	\label{eq:DCstandard}
	\frac{\dd D}{\dd f}=\int \dd z \frac{\dd R}{\dd z} \frac{\dd \bar\tau}{\dd f}.
\end{equation}
In the equation above, the event rate per redshift slice $\dd R/\dd z$ is given in terms of the comoving distance  $r=\int^z_0 \frac{\dd z^\prime}{H(z^\prime)}$ and the merger rate as
\begin{equation}
	\frac{\dd R}{\dd z} = 
	\frac{1}{(1+z)}\frac{4\pi r^2}{H}
	\int \dd m_1\,
	\dd m_2 \,
	\frac{\dd^2 \tau_{\rm merg}(z,\,m_1,\,m_2)}{\dd m_1 \dd m_2},
	\label{eq:dRdz_general}
\end{equation}
and the duration of the signal at frequency $f$ can be written in terms of the chirp mass as
\begin{equation}
	\frac{\dd \bar\tau}{\dd f}  = \frac{5}{96\pi^{8/3}}(G{\cal M}_c^z)^{-5/3}f^{-11/3},
\end{equation}
and is also integrated over $\dd m_1\,\dd m_2$ together with the integrand in Eq.~\eqref{eq:dRdz_general}.
The total duty cycle in the frequency range $[f_{\rm min}, \,f_{\rm max}]$ is simply obtained by integrating over frequencies as
\begin{equation}
	\xi=\int_{f_{\rm min}}^{ f_{\rm max}} \dd f \frac{\dd D}{\dd f} \,.
\end{equation}

With the help of the duty-cycle, we are now in the position to define the popcorn and continuous contributions to the SGWB as follows~\citep{Regimbau:2011bm}:
\begin{equation}
	\Omega_{\rm GW}(f)=	\Omega_{\rm GW}^{\rm pop}(f) + \Omega_{\rm GW}^{\rm cont}(f),
	\label{Eq:OGWseparation1}
\end{equation}
where
\begin{align}
	&\Omega_{\rm GW}^{\rm pop}(f)\equiv\Omega_{\rm GW}(f, \,0,\,z^*(f))\,,\\
	&\Omega_{\rm GW}^{\rm cont}(f)\equiv\Omega_{\rm GW}(f, \,z^*(f),\,z_{\rm max})\,,
	\label{Eq:OGWseparation2}
\end{align}
and, at given frequency $f$, the function $z_*(f)$ is defined as the redshift that solves $dD/df(f)=N$. The authors of~\citep{Coward:2006df,Regimbau:2011bm} use $N=10$ to define the boundary between continuous and popcorn background whereas a less stringent $N=1$ is used in \citep{Damour:2000wa,Damour:2004kw}. We adopt the latter in the following, which physically means that the time interval between two consecutive GW events is the same as the duration of the signal emitted in the frequency bin  $[f,\,f+ \dd f]$. In other words, $N=1$ corresponds to the situation where we always find one GW event on average in the frequency bin.

The continuous background consists of overlapping signals at a given frequency, either because the number of sources is very large or because the signal duration is long compared to the time between consecutive events. Because of the central limit theorem, the background is well described by Gaussian statistics, implying that the optimal analysis method is the cross-correlation statistic \citep{Allen:1997ad}. On the other hand, the popcorn background is strongly non-Gaussian and arises from unresolvable signals whose duration is comparable to the distance between consecutive events, in a way that the signal at the given frequency is not continuously present. Because of its non-Gaussian properties, data analysis techniques beyond cross-correlation need to be adopted in searches of popcorn backgrounds ~\citep{Drasco:2002yd,Seto:2008xr,Seto:2009ju,Thrane:2013kb,Martellini:2014xia,Martellini:2015mfr,Cornish:2015pda,Smith:2017vfk,Smith:2020lkj,Yamamoto:2022kuh}.

\subsection{Definition with the horizon distance}
\label{detectorDC}

The standard expression of the duty cycle, Eq.~\eqref{eq:DCstandard}, only depends on the properties of the source population, and it is obtained by integrating the source redshift from zero to infinity. This quantity is useful to describe whether the SGWB is in the continuous or popcorn regime in general discussion. 
However, 
when we consider real GW data of a popcorn background, very low SNR ($\ll 1$) signals would be lost among the detector noise and would not contribute as popcorn. Such events should be removed from the computation when we want to make a theoretical prediction for the purpose of comparing it with an observed value of the duty cycle, possibly achievable by future experiments. Note that this argument applies only when events are sparse and the low SNR signal is isolated from the other events. When low SNR events are numerous, their accumulation could form a continuous SGWB with a detectable amplitude. In that case, we should not remove low SNR events as they are making an essential contribution, although in that case, we have $dD/df \gg 1$ and the value of the duty cycle no longer provides useful information on the statistical property as it just becomes Gaussian. We will make a special note for such a case.  

We now derive a detector-dependent duty cycle by replacing the upper limit of the redshift integration with the horizon distance of the detector. This quantity is clearly not an intrinsic property of the background itself, but it rather depends on the detector sensitivity and can be used to compare with future data analysis. As noted above, this detector-dependent duty cycle would provide useful information only for $dD/df\ll 1$.

The horizon distance of the detector at a given frequency $f$ depends on the chirp mass and is given by~\citep{VIRGO:2012kov,Chen:2017wpg,Carr:2019kxo}
\begin{align}
	R_{\rm det}(f,\,m_1,\,m_2,\,z)=&\frac{c \pi^{-2/3}}{{\rm SNR_{th}}\times2.26}\,
    \sqrt{\frac{5}{6}}
    \left(\frac{G\mathcal{M}_c^z}{c^3}\right)^{5/6}\notag\\\label{eq:Rdet}
	&\times\left(\int_{f_{\rm min,\, det}}^{f_{\rm max}(f)}df'\,\frac{f'^{-7/3}}{S_h(f')}\right)^{1/2},
\end{align}
where the detector sensitivity is in the range $[f_{\rm min,\, det},\,f_{\rm max,\, det}]$ and
\begin{align}
	f_{\rm max}(f)\equiv&\max\left(f_{\rm min,\, det},\,\min[f,\,\min\left(f_{\rm max,\, det},\,f_*\right)] \right),\notag\\
	f_*\equiv &2f_{\rm ISCO} \frac{4 m_1 m_2}{(m_1+m_2)^2}\,,\notag\\
	f_{\rm ISCO}\equiv& 4.4\,{\rm kHz}\frac{M_\odot}{m_1+m_2} \,.
\end{align}

The horizon distance represents the distance of the furthest detectable source with a signal-to-noise ratio exceeding a given value of SNR.

The generalization of Eq.~\eqref{eq:DCstandard} that includes the sensitivity of the horizon is straightforward. For each frequency $f$ and masses $(m_1,\,m_2)$, we solve the equation $R_{\rm det}(f,\,m_1,\,m_2,\,z_{\rm det})=d_L(z_{\rm det})$ and obtain the redshift corresponding to the horizon distance of the detector $z_{\rm det}=z_{\rm det}(m_1,\,m_2,\,f)$. The new expression for the {\em detector-dependent} duty cycle is:
\begin{equation}
	\label{eq:DC_DET}
	\xi_{\rm det}=\int_{f_{\rm min}}^{ f_{\rm max}} \dd f \frac{\dd D_{\rm det}}{\dd f}=\int_{f_{\rm min}}^{ f_{\rm max}} \dd f \int_0^{z_{\rm up}} \dd z \frac{\dd R}{\dd z} \frac{\dd \bar\tau}{\dd f},
\end{equation}
where now 
\begin{equation}
	\label{eq:zUp}
	z_{\rm up}= {\rm Min}\Bigg[ z_{\rm max}(m_1,\,m_2,\,f),\, z_{\rm det}(m_1,\,m_2) \Bigg] \,.
\end{equation}
Note that, by definition, we always have $\xi_{\rm det}\leq\xi$.

\section{Results}
\label{sec:results}

\subsection{Benchmark models}
\label{sec:result1}

\begin{figure*}
	\includegraphics[width=\columnwidth]{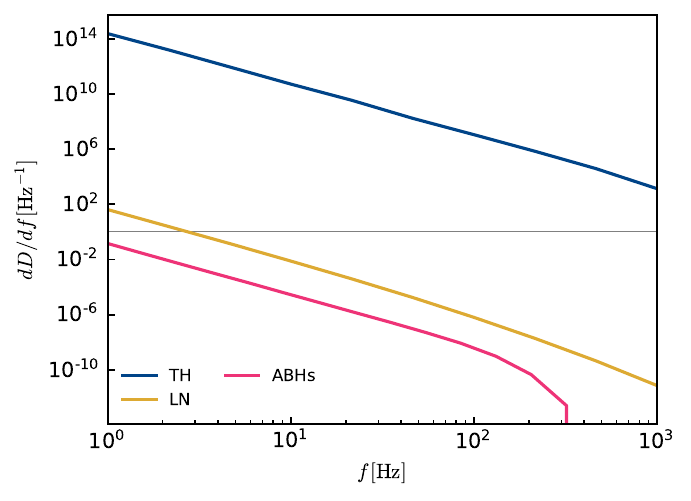}
	\includegraphics[width=\columnwidth]{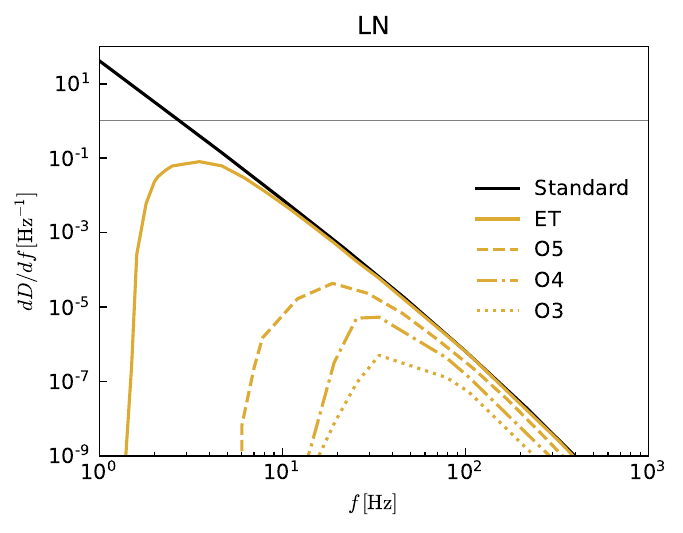}
	\includegraphics[width=\columnwidth]{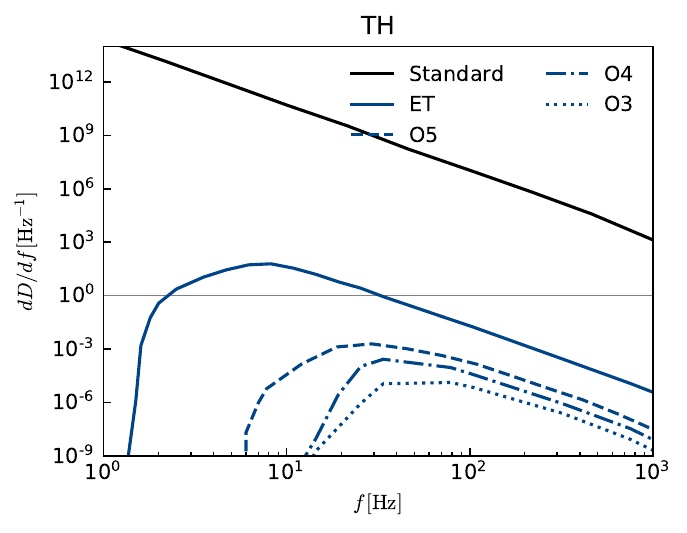}	\includegraphics[width=\columnwidth]{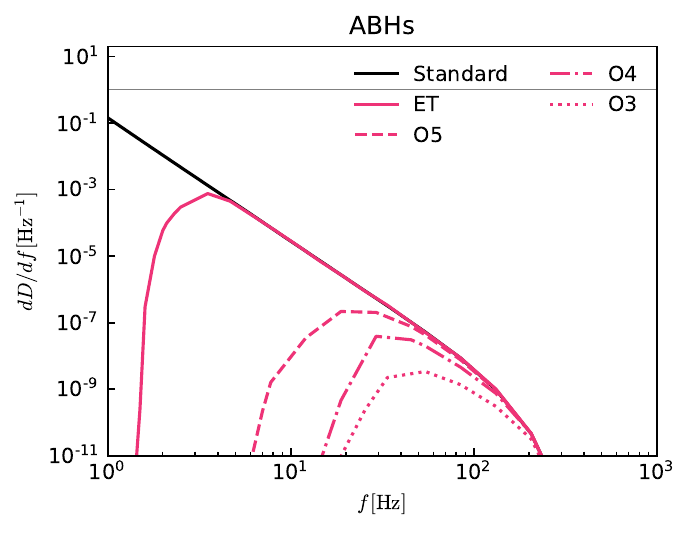}
	\caption{\label{fig:DC} [Top-Left] Differential duty cycle for three different models (LN and TH models for PBH, compared to the ABH case). We also plot the detector-dependent differential duty cycle for the LN model [Top-Right], for the TH model [Bottom-Left] and for ABHs [Bottom-Right]. We assume the baseline parameters described in Sec.~\ref{sec:mass_function} for all of the models.}
\end{figure*}

\begin{table}
	\centering
	\caption{Total duty cycle $\xi_{\rm det}$ for different models and detectors. To get $\xi_{\rm det}$, we integrate the differential duty cycle $dD_{\rm det}/df$ over the frequency range $[2,\,1000]$ Hz for ET and $[10,\,1000]$ Hz for LIGO O3, O4, and Advanced-LIGO.}
	\label{tab:dc}
	\begin{tabular}{|c||c|c|c|c|c||}
		\hline
		& O3 & O4    & AdvLIGO & ET \\ \hline  
		LN& $1.6\times10^{-5}$     & $10^{-4}$  & $6\times10^{-4}$   & $0.33$ \\ \hline
		TH  & $1.7\times10^{-3}$ & $0.02$   & $0.08$ & $532.6$\\ \hline
		ABHs  & $2.0\times10^{-7}$   &  $1.5\times10^{-6}$  & $7.0\times10^{-6}$ &  $2.2\times10^{-3}$\\ \hline
	\end{tabular}
\end{table}

\begin{figure}
	\includegraphics[width=\columnwidth]{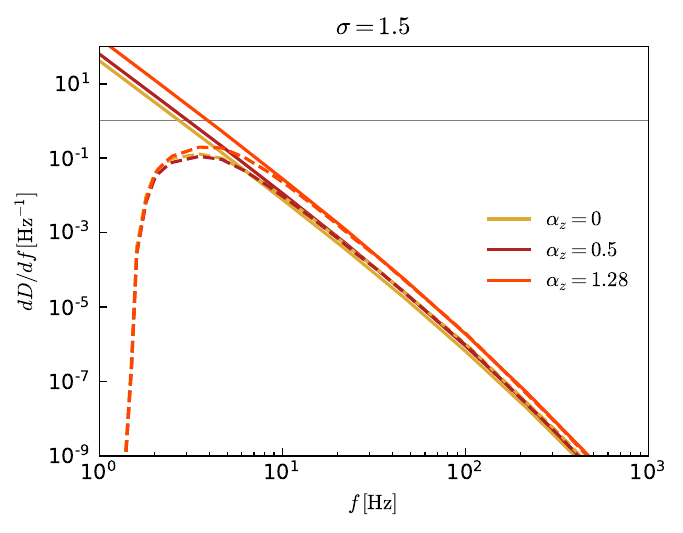}
	\includegraphics[width=\columnwidth]{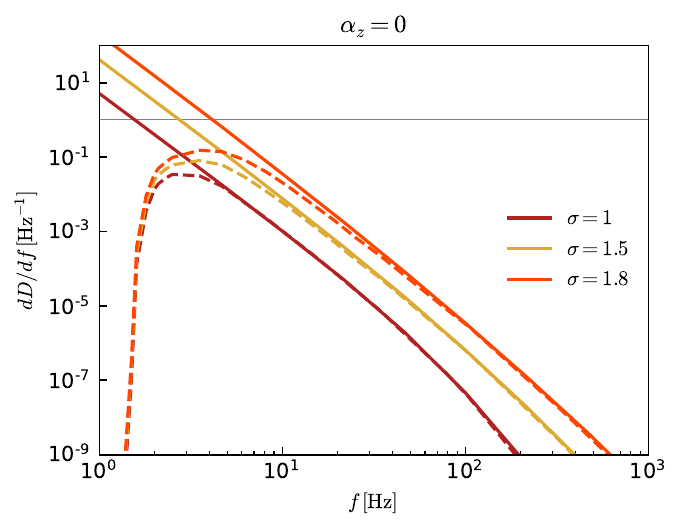}
	\includegraphics[width=\columnwidth]{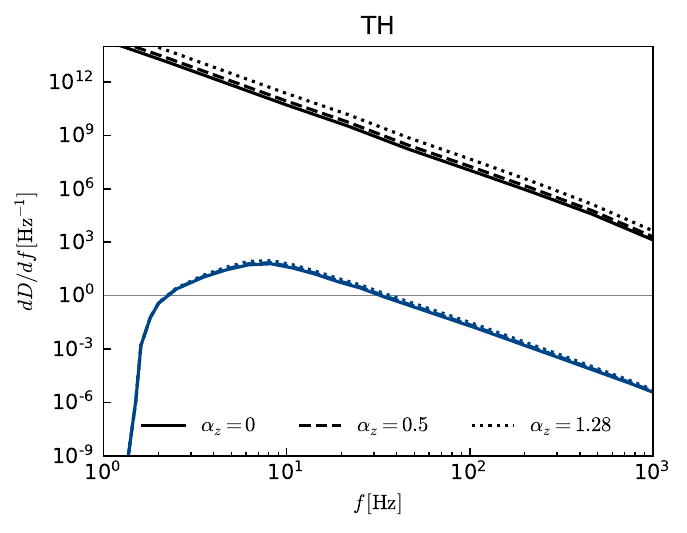}
	\caption{\label{fig:dc_det_variation} Duty cycle for different PBH model parameters. We vary the redshit dependence of the merger rate, parameterized by $\alpha_z$,  in the LN and TH  models in the top and bottom panels respectively. In the center panel, we vary the width of the LN mass function. }
\end{figure}

Having introduced the necessary theoretical formalism, we now investigate the detection regimes of the  BH populations introduced above. In Fig.~\ref{fig:DC}, we show their duty cycle for the baseline models described in Section~\ref{sec:mass_function}.

Let us start by commenting on the top-left panel, which compares the detector-independent duty cycle of the three different models (PBHs with LN and TH mass functions, and ABHs). Interestingly, both PBH models predict the duty cycle larger than that of ABHs. This remains the same even for the detector-dependent duty cycle, as can be observed in the other panels, where we fix ${\rm SNR_{th}}=1$ in Eq.~\eqref{eq:Rdet} and compute $\dd D_{\rm det}/\dd f$ for the O3, O4, Advanced-LIGO design sensitivity, and ET. 

As anticipated in the previous sections, there are two reasons why this happens. The first is the different redshift dependence of the merger rate between PBHs and ABHs (see right panel of Fig.~\ref{fig:tau}). Since Eqs.~\eqref{eq:DCstandard} and ~\eqref{eq:DC_DET} are essentially integrals  of the merger rate over redshift, many distant PBH events with a very small ${\rm SNR}$ contribute to the integral, yielding a large $\xi$ and $\xi_{\rm det}$. As explicitly reported in Table~\ref{tab:dc}, the duty cycle of PBHs is generically two orders of magnitude larger than that of ABHs.
The second reason is that the duty cycle crucially depends on the mass function of the BH population. This is particularly relevant for the TH PBH model, which is described by a very broad mass function. The model thus predicts many events with several combinations of the total mass and mass ratio, but the largest contribution comes from $2\,M_\odot$ BHs. Compared to the astrophysical case or the LN case with a peak at $30\,M_\odot$, such small mass BHs typically create events with a longer duration, thus resulting to larger $\dd D/\dd f$.

Note that, in Table~\ref{tab:dc}, we show the detector-dependent duty cycle for all models. For the TH model, we see from Fig.~\ref{fig:DC} that $\xi\gg 1$ and the SGWB is continuous, and we should look at the detector-independent one $\xi$. However, we still show the values of $\xi_{\rm det}$ since it can be useful information for analyses like \citep{Smith:2020lkj} to know the number of sub-threshold events in each experiment.
	
In Fig.~\ref{fig:dc_det_variation}, we show the results by changing  the different redshift dependence of the merger rate $\alpha_z$. As expected, the duty cycle increases for larger values of $\alpha_z$, which corresponds to the growth of the PBH merger rate with redshift. Although it is not easy to appreciate it from the figures plotted in log-scale, the difference in the duty cycle is of $\mathcal{O}(1)$. For the LN model, we also show the effect of increasing or decreasing the width of the LN mass function.  Higher values of  $\sigma$ increase the duty cycle since a broader tail of the mass function leads to more events with small mass, with large event duration. The variation is more significant than the one for $\alpha_z$.

\begin{figure*}
	\includegraphics[width=\columnwidth]{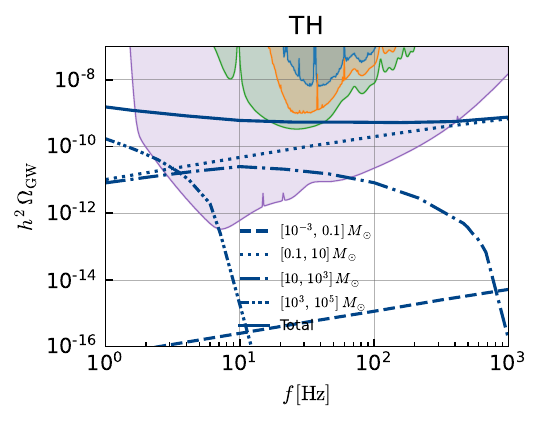}
	\includegraphics[width=\columnwidth]{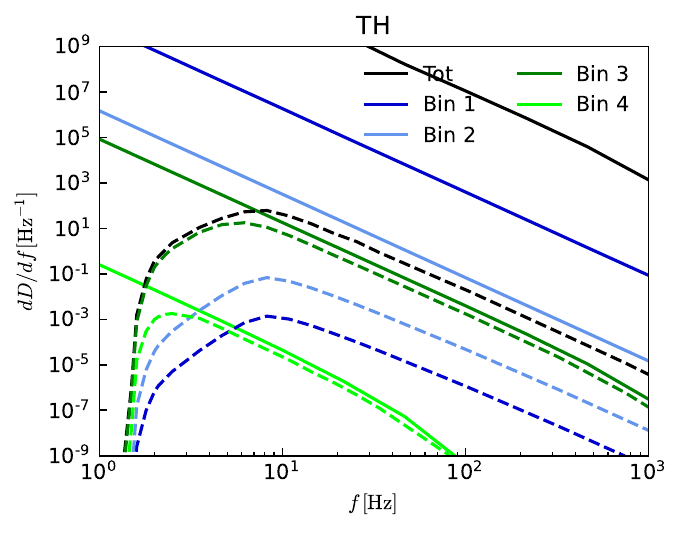}	\includegraphics[width=\columnwidth]{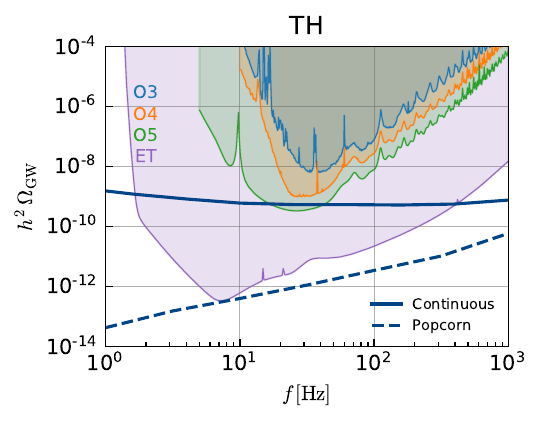}
	\includegraphics[width=\columnwidth]{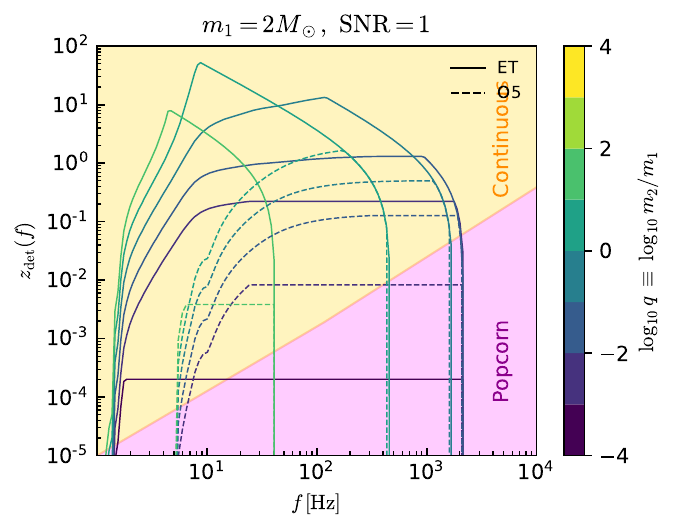}
	\caption{\label{fig:broad_mf} Figures to show some details about the TH model. [Top-Left] Contribution of different mass bins to the energy density of the SGWB. [Top-Right] Contribution of different mass bins to the duty cycle. Solid and dashed lines represent the detector-independent and detector-dependent duty cycle (ET is assumed), respectively. The mass bins are divided as follows: Bin 1 $=[10^{-5},\,10^{-3}]\,M_\odot$, Bin 2 $=[10^{-3},\,10^{-1}]\,M_\odot$, Bin 3 $=[10^{-1},\,10]\,M_\odot$ and Bin 4 $=[10,\,10^{3}]\,M_\odot$. [Bottom-Left] Comparison of the continuous (solid) and popcorn (dashed) contributions, separated by the threshold $dD/df = 1$. [Bottom-Right] We plot $z_{\rm det}(f)$ for different values of the mass ratio $q\equiv m_2/2 M_\odot$ (solid curves for ET and dashed curves for Advanced-LIGO) on top of the redshift bands separated by the continuous and popcorn regimes.}
\end{figure*}

As the TH model shows quite a different behavior compared to the other models, let us look into some details with the help of Fig.~\ref{fig:broad_mf}. The first two figures show how different mass bins contribute to the SGWB amplitude and the duty cycle. The top-left panel is obtained by integrating Eq.~\eqref{Eq:OGW} over smaller mass bins. Note that the sum of $\Omega_{\rm  GW}$ for each mass bin is not equal to $\Omega_{\rm GW}$ integrated over the full range (solid line) because mergers with large mass ratio (i.e. BHs paired beyond the mass bins) are not included, but the figure is useful for illustrative purposes. 
As seen in the figure, the mass bin of $[0.1,\,10]\,M_\odot$ is giving the dominant contribution on the SGWB amplitude, and they are essentially in their inspiral phase as we see the $f^{2/3}$ dependence of the spectrum.
The top-right panel of Fig.~\ref{fig:broad_mf} shows the duty cycle calculated for different mass bins. As can be seen, the mass bin of $[0.1,\,10]\,M_\odot$  is giving the dominant contribution also to the duty cycle. Many of them have a very small ${\rm SNR}$, but they sum up as incoherent noise in the detector and produce continuous SGWB, dominating the popcorn component. This can be seen in the bottom-left panel of Fig.~\ref{fig:broad_mf}, which is produced by using Eqs.~\eqref{Eq:OGWseparation1} -- \eqref{Eq:OGWseparation2}. As mentioned in Sec.~\ref{detectorDC}, in such a case, we should look at the detector-independent duty cycle, but in the figure, we also plot the detector-dependent duty cycle, which helps to see that most of the events indeed have ${\rm SNR}\ll 1$ and each event cannot be resolved as a single event.

 \begin{figure*}
	\includegraphics[width=\columnwidth]{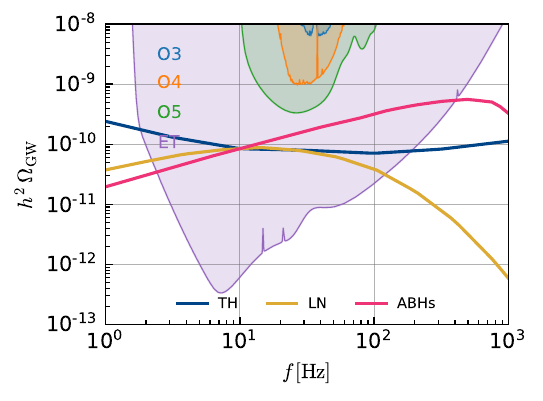}	\includegraphics[width=\columnwidth]{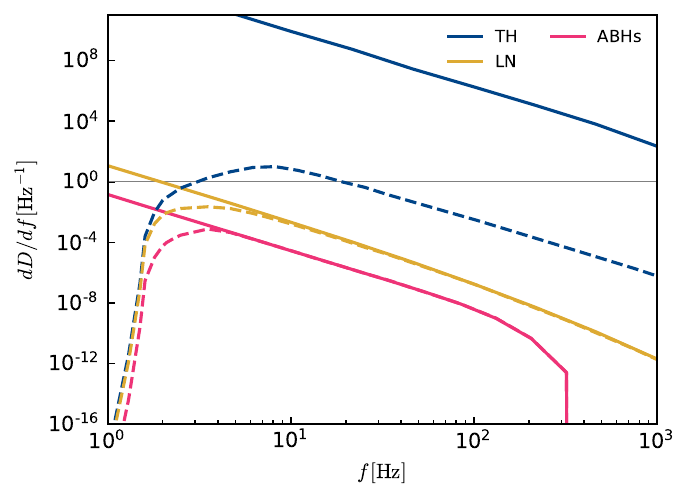}
	\caption{\label{fig:Omega_amplitude} [Left] $\Omega_{\rm GW}$ normalized to the same amplitude. [Right] Detector-independent (solid) and dependent (dashed) duty-cycle corresponding to the backgrounds shown in the left panel.}
\end{figure*}

Let us now comment on the bottom-right panel of Fig.~\ref{fig:broad_mf}. We show the redshift corresponding to the horizon distance of the detector $z_{\rm det}(f,\,m_1=2\,M_\odot,\,m_2=q\,m_1)$ for both ET (solid lines) and AdvLIGO (dashed lines). Together, we also show the popcorn and continuous regimes using a threshold of $dD/df = 1$. As can be seen, only very nearby binaries contribute to the popcorn regime and a very large number of them from many decades in redshift emit signals that superimpose to create a continuous background.

Finally, let us note that in the last part of this Section, we focused on the TH model because the differences with respect to ABHs are maximized in this model. However, the statistical properties of LN PBHs are also different as their duty cycle differs by two orders of magnitude. The next generation of ground-based detectors should be sensitive to such a large difference  
~\citep{Smith:2020lkj}, suggesting that the duty cycle could be used to distinguish the primordial and astrophysical nature of BHs, as we elaborate in the next Subsection.

\subsection{Distinguishing between PBHs and ABHs using the duty cycle}

In the previous subsection, we normalized the amplitude of the SGWB for each model by fixing the merger rate to explain the rate of the observed BBHs in the GWTC-3 catalog. Here, we consider a different setup. Suppose that ET telescope measures a SGWB, but different models can explain the observed amplitude of SGWB as shown in the left panel of Fig.~\ref{fig:Omega_amplitude}. The question we would like to address is: can we use its statistical properties, i.e., the duty cycle, to infer which sources produced it? This information would be highly complementary to the spectral shape of the SGWB.

For this purpose, in this subsection, we normalize the amplitude of the SGWB for all the models to be the same at $f=10$ Hz, roughly the frequency of the best sensitivity of ET. In this case, the merger rate of PBHs is lower than the one used in the previous subsection and cannot explain the merger rate of observed BBHs. Thus, we consider the situation where the LVK rate is essentially explained by ABHs and PBHs only give a subdominant contribution. We note that the normalization of PBH merger rate has nothing to do with the fraction of CDM as our merger rate model described in Eq.~\eqref{eq:tauz} depends on the combination $R_{\rm clust} f_{\rm PBH}^{\rm tot\,\,2}$, which is the known degeneracy between clustering and abundance of PBHs~\citep{Raidal:2017mfl,Clesse:2020ghq,Vaskonen:2019jpv,Young:2019gfc,Trashorras:2020mwn,Atal:2020igj,DeLuca:2020jug}. 

The plot on the right panel of Fig.~\ref{fig:Omega_amplitude} shows that, indeed, we could pin down the source of the background by looking at its duty cycle. We see that, even if the backgrounds produced from PBHs have amplitudes smaller than those in Fig.~\ref{fig:omega}, as we lowered $\Omega_{\rm GW}$ to match the one from ABHs at $f=10$ Hz, their duty cycle is still larger than that of ABHs, and we find an order of magnitude difference between the total duty cycle of ABHs ($\xi_{\rm ET}=2.2\times10^{-3}$) and of LN PBHs ($\xi_{\rm ET}=0.10$). Previous studies~\citep{Smith:2017vfk,Yamamoto:2022kuh} have simulated the parameter estimation of the non-Gaussian background and showed that the error on $\xi$ could be $10-20\%$ if the background is detected with a certain SNR. Thanks to the large difference between the scenarios scenarios we considered, even if we take into account uncertainties that could affect the estimation of the duty cycle, such as glitch noise, we could still expect the next generation of ground-based detectors to tell apart the astrophysical and primordial channels for the formation of BHs.

We end by noting that in the case of the TH model, we have $\dd D/\dd f\gg1$, which indicates that the background is always in the continuous regimes and, restricting to this model, standard cross-correlation searches for the SGWB are already adequate to place constraints on this specific model. LN PBHs, on the other hand, may require adopting more sophisticated analysis techniques to take into account their popcorn nature correctly.

\section{Conclusions}
\label{sec:conclusions}

In this paper, we have studied the popcorn signature as a tool to understand the origin of the SGWB produced by unresolved BBHs at the frequency of ground-based interferometers. A popcorn background is characterized by GW signals that may or may not overlap in the frequency band of the detector, unlike a continuous background where the sensitivity band is constantly occupied by the superposition of GW signals which effectively act as Gaussian noise. In order to characterize the two regimes of the background, we have made use of the duty cycle, which represents the average number of events present in a given frequency band, and is larger (smaller) than one for a continuous (popcorn) background. 
	
Our main finding is that populations of binaries formed from PBHs predict a duty cycle that is generically orders of magnitude larger than the one for astrophysical binaries. This opens up the possibility to discriminate between astrophysical and primordial BHs using the statistical properties of the SGWB. Our results are especially relevant to population analyses of sub-threshold events. 

To this purpose, we have also proposed a new way to compute the duty cycle, which takes into account the sensitivity of the detector. Our procedure selects only the events contributing to the duty cycle with SNR larger than a certain threshold, which we take as ${\rm SNR}=1$. 
This quantity would be useful when comparing the theoretical prediction with an observed duty cycle. Note that the value of the SNR thres\-hold should be determined by how much the data analysis can be sensitive to sub-threshold events and how it affects the measurement of the duty cycle. It should be investigated in more detail.

We have shown that each model predicts different values of the duty cycle. Our results indicate that not only are ABHs distinguishable from primordial ones based on their duty cycle, but we can also tell apart different PBH models. In fact, while the LN model (or any other PBH model with a mass function peaked around $\mathcal{O}(10)\,M_\odot$) is characterized by popcorn signatures that differ from ABHs, the TH model, whose mass function is very broad, does not show popcorn signatures at all. Rather, the background they produce is continuous. The peculiar spectral shape of the SGWB, almost flat in the $[1,\,1000]\,{\rm Hz}$ range, is yet another observable to distinguish PBHs with very broad mass functions to LN PBHs and astrophysical ones.

We also would like to point out an interesting application of the duty cycle. It can also be used to estimate the number of unresolved multiple simultaneous events in the LVK band. Especially, for the TH model, $2\,M_\odot$ PBHs merge at high frequencies where LVK has a poor sensitivity due to quantum shot noise, and they leave only a signal of the inspiral phase lasting a few minutes inside the detector sensitivity, making it difficult to distinguish them from each other. If their merger rate is high enough, as in the TH model, such events may overlap with other individual events and cause wrong parameter estimation. Eventually, with the much better sensitivity of the Einstein Telescope, we may be able to separate the different components into individual events, but otherwise, sub-threshold events could act as correlated non-Gaussian noise. So, even before the detection of the SGWB, the first indication of the large peak at $\sim 2\,M_\odot$ in the TH mass function could be the presence of multiple simultaneous sub-threshold events. The values of the detector-dependent duty cycle calculated in this paper can be used to estimate the probability of such events.
	
Although our results are very optimistic and show the promising prospects of testing PBHs with the popcorn signature, there are some caveats in our analysis related to the modeling of the merger rate in the clustered PBHs scenario. Indeed, the clustering of PBHs enhances both the probability of forming binaries, but also the one of disrupting them before their merger~\citep{Raidal:2018bbj,Trashorras:2020mwn}, which is not taken into account in our analysis. Another issue would be to include the early binary formation channel with a better understanding of the effect of a broad mass function.
 In that case, supposing that we normalize the local merger rate to be $45\,{\rm yr}^{-1}\,{\rm Gpc}^{-3}$, as done in Sec.~\ref{sec:result1}, the difference appears only in the mass dependence of the merger rate. As we have seen in the comparison between the LN and TH models, the major factor affecting the value of the duty cycle is the difference in the mass function. Also, we have seen that the effect of the difference in the merger rate evolution is relatively small. Therefore, we do not expect the duty cycle to change significantly, even if we change the merger rate model to the early binary formation scenario. This should be true in the case of the LN mass function, which is very peaked, and the mass dependence of the merger rate becomes less important. On the other hand, the TH mass function has a broad mass spectrum, and it may cause a non-trivial difference in the value of the duty cycle. However, the duty cycle in this model is much larger than unity, so we do not expect this to change the conclusion that the stochastic background is continuous for the TH model.

For both PBHs and ABHs, the details for the merger rate estimation still need to be thoroughly investigated and should be prepared before the detection of the SGWB. Nevertheless, this work has provided an important step toward using the duty cycle as a new observable to discriminate the primordial and astrophysical origin of the SGWB.

\section*{Acknowledgements}

This work is supported by the Spanish Research Projects PGC2018-094773-B-C32 (MINECO) and PID2021-123012NB-C43 (MICINN-FEDER) and the Centro de Excelencia Severo Ochoa Program CEX2020-001007-S. MB and SK are supported by the Spanish Atracci\'on de Talento contract no. 2019-T1/TIC-13177 granted by Comunidad de Madrid, the I+D grant PID2020-118159GA-C42 of the Spanish Ministry of Science and Innovation and the i-LINK 2021 grant LINKA20416 of CSIC. SK is partially supported by Japan Society for the Promotion of Science (JSPS) KAKENHI Grant no. 20H01899 and 20H05853. 
\vspace{0.5cm}

\section*{Data Availability}
The data underlying this article will be shared on reasonable request
to the corresponding author.
\vspace{0.5cm}

%%%%%%%%%%%%%%%%%%%%%%%%%%%%%%%%%%%%%%%%%%%%%%%%%%
%\section*{Data Availability}

%The inclusion of a Data Availability Statement is a requirement for articles published in MNRAS. Data Availability Statements provide a standardised format for readers to understand the availability of data underlying the research results described in the article. The statement may refer to original data generated in the course of the study or to third-party data analysed in the article. The statement should describe and provide means of access, where possible, by linking to the data or providing the required accession numbers for the relevant databases or DOIs.



\bibliographystyle{mnras}
\bibliography{example} % if your bibtex file is called example.bib

% This method is tedious and prone to error if you have lo\ddts of references
%\begin{thebibliography}{99}
%\end{thebibliography}

%%%%%%%%%%%%%%%%%%%%%%%%%%%%%%%%%%%%%%%%%%%%%%%%%%

%%%%%%%%%%%%%%%%% APPENDICES %%%%%%%%%%%%%%%%%%%%%

\noindent
%%%%%%%%%%%%%%%%%%%%%%%%%%%%%%%%%%%%%%%%%%%%%%%%%%

% Don't change these lines
\bsp	% typesetting comment
\label{lastpage}
\end{document}